\documentclass[reprint,amsmath,amssymb,aps,prx,superscriptaddress]{revtex4-2}
\usepackage{graphicx}
\usepackage{dcolumn}
\usepackage{bm}

\providecommand{\aap}{Astronomy \& Astrophysics}
\providecommand{\aj}{The Astronomical Journal}
\providecommand{\apj}{The Astrophysical Journal}
\providecommand{\apjl}{The Astrophysical Journal Letters}
\providecommand{\apjs}{The Astrophysical Journal Supplement Series}
\providecommand{\mnras}{Monthly Notices of the Royal Astronomical Society}

\providecommand{\nat}{Nature}
\providecommand{\physrep}{Physics Reports}

\providecommand{\physscr}{Physica Scripta}
\begin{document}

\preprint{APS/123-QED}

\title{Tracing the evolution of galaxy environments with manifold learning}

\author{Ana Sof\'ia M. Uzsoy}
 \email{ana\_sofia.uzsoy@cfa.harvard.edu}
\affiliation{%
 Center for Astrophysics $|$ Harvard \& Smithsonian, Cambridge, MA}

\author{Claire Lamman}
\affiliation{
 Center for Cosmology and AstroParticle Physics (CCAPP), The Ohio State University, Columbus, OH
}%
\affiliation{
 Department of Astronomy, The Ohio State University, Columbus, OH
}

\author{Peixin Zhu}
\affiliation{%
 Center for Astrophysics $|$ Harvard \& Smithsonian, Cambridge, MA}
\date{\today}
\author{Melanie Weber}
\affiliation{%
John A. Paulson School of Engineering and Applied Sciences, Harvard University, Cambridge, MA}%

\begin{abstract}
We present {\tt SONDE} (SOrted Neighbor Distance Embedding): a scalable manifold learning approach to create a nuanced, continuous, low-dimensional representation of the geometry of galaxies' local environments. {\tt SONDE} is a variation on Isomap that uses similarity distances between galaxy neighborhoods to create an embedding space, within which galaxies cluster according to their neighborhood geometries. We additionally present a novel method to project different datasets into the same embedding space, enabling comparisons of environmental embeddings across redshifts. We demonstrate this method on the TNG100 simulations at $z = 0 - 5$ and show that the embedding values evolve with redshift and correlate with galaxies' physical properties. We train a neural network to predict physical properties based on environmental features and show that our first embedding dimension outperforms a standard environmental quantifier, the number of neighbors, while the top three dimensions encode as much environmental information as a ten-bin radial profile. Our results are consistent with known density correlations, including galaxy halo mass, central vs. satellite status, and quenching.  {\tt SONDE} provides a straightforward way to encode subtleties of galaxy environments that could have significant implications for future analysis of galaxy-environment interactions.
\end{abstract}

\maketitle


\section{Introduction}

The distribution of galaxies throughout the Universe traces the underlying large-scale structure known as the ``cosmic web'' \cite{1996Natur.380..603B}. 
The cosmic web evolves over time, driven by gravitational instability and perturbations in the Universe's dark matter distribution \cite{zeldovich}. Previous work in simulations show that the cosmic web evolves through hierarchical structure formation, with larger, more complex structures being formed over time \cite{2014MNRAS.441.2923C}.
Characterizing substructures within the cosmic web and quantifying how they evolve is a fundamental step towards understanding how galaxy properties are influenced by their environments at different times in the Universe's history.

Various discrete methods exist to classify substructures of galaxies in the cosmic web \cite{2018MNRAS.473.1195L}. Clustering algorithms such as k-nearest neighbors \cite{knn}, DBSCAN \cite{dbscan}, and friends-of-friends \cite{fof, 2011ApJS..195....4M} assign galaxies a binary label of if they are in an identified cluster or not. One challenge with these methods is that they enforce strict cluster boundaries, when a cluster's environmental influence can extend beyond its immediate radius \cite{hao_finding_2025}. Topological data analysis can identify filaments, walls, voids, and other substructures in distributions of galaxies \cite{2011MNRAS.414..350S, 2011MNRAS.414..384S, 2017MNRAS.465.4281P, 2021MNRAS.507.2968W}, but these are also generally discrete classifications.

Continuous methods allow for more nuanced characterization of the cosmic web. Local density estimation, denoting how many neighboring galaxies are within a certain radius, is one of the most prominent continuous metrics of a galaxy's local environment. The Hessian of the smoothed density field allows for classification of voids, filaments, walls, and nodes based on its eigenvalues \cite{2014MNRAS.441.2923C}. Two-point clustering statistics quantify the number of galaxy pairs within a certain distance that are in excess of a random distribution, and is a standard metric for quantifying the structural nature of a distribution of points \cite{3e7eb752b0684da2a188d4b6f73e3f1c}. Higher-order clustering can probe this even further; the cumulative distribution function of the distances between random points and galaxies can provide a scalable way of probing higher-order information \cite{2021MNRAS.500.5479B}.

One of the most comprehensive methods for characterizing the cosmic web is with the tidal tensor, the Hessian of the gravitational potential, which captures higher-order information about the curvature of the surface at a given point. Individual substructures can then be identified by the sign of the eigenvalues of the tidal tensor denoting clusters, filaments, sheets, and voids \cite{hessian_cosmic_web, 2009MNRAS.396.1815F}. This classification can be accelerated using graph neural networks \cite{kololgi_learning_2026} or diffusion models \cite{noor_learning_2026}, and the ratios of these eigenvalues can be used as a continuous metric for the structure of the cosmic web \cite{eigenvalue_ratios}.  Constructing a continuous potential field and calculating the Hessian for every galaxy in a large dataset can be computationally expensive, and the method of classifying based on eigenvalues imposes a predetermined set of structures.

Manifold learning is a subfield of geometric machine learning that involves finding informative, low-dimensional representations of complex datasets. The crux of manifold learning is using information about intrinsic geometry, symmetry, or equivariance of the dataset to optimize dimensionality reduction. Common manifold learning methods include Isomap \cite{isomap}, Locally Linear Embeddings \cite{lle}, and Laplacian Eigenmaps \cite{laplacian_eigenmaps}, each of which take a slightly different approach to representing a high-dimensional dataset on a low-dimensional manifold.

In this work, we present {\tt SONDE} (SOrted Neighbor Distance Embedding), a new manifold learning method that encodes galaxy environments in a continuous latent space based on the geometry of their surrounding neighborhoods. We additionally present a novel approach for projecting different datasets into a shared latent space that enable a detailed analysis of how the geometry of galaxy environments has evolved from 12.6 billion years ago ($z = 5$) to the present day ($z = 0$) in the TNG100 simulations. Our unsupervised method produces a flexible, continuous embedding space that can encode a wide variety of environments instead of classifying galaxies into a discrete, predetermined set of structures, making it particularly well-suited for exploratory data analysis.

This paper is structured as follows:
in Section~\ref{sec:method} we provide a detailed description of {\tt SONDE} for embedding one dataset of galaxy positions and for aligning embedding spaces across different redshift snapshots so that their embeddings can be compared. We additionally discuss the assumptions made by our method and the implementation details. In Section~\ref{sec:results} we show the results of our method first on the $z = 0$ snapshot, interpreting the top embedding dimensions using two-point clustering and correlation coefficients, before exploring how the neighborhood embeddings and physical properties of galaxies evolve across $z = 0 - 5$. We additionally use a neural network to predict physical properties based on the neighborhood embeddings and compare this to the predictive power of the number of neighbors. We further discuss the implications and limitations of {\tt SONDE} in Section~\ref{sec:discussion} before concluding in Section~\ref{sec:conclusions}.

\section{Method Overview}\label{sec:method}
Our aim is to create a low-dimensional embedding space where galaxies cluster based on their local neighborhood geometry. Here we use the term ``neighborhood'' to mean the radial distribution of surrounding galaxies and do not directly encode any information about the underlying dark matter halo.

The input to our method is an $n \times 3$ matrix consisting of $n$ points and their 3D spatial coordinates, and the output is a $n \times m$ matrix containing representations of each of the $n$ points in an $m$-dimensional embedding space, where $m << n$. The hyperparameters of this method are the local neighborhood radius $r$, which denotes the scale of the surrounding neighborhood that is to be considered, the inverse distance cutoff $c$, where $1/c$ is the smallest included distance to a neighbor, and the dimensionality $m$ of the embedding space \footnote{A preliminary version of this work was presented at the Machine Learning and the Physical Sciences Workshop @ NeurIPS 2025 \cite{2025manifolds}}. 

\subsection{Embedding a single dataset}

We first calculate pairwise distances between galaxies within the neighborhood radius $r$, using periodic boundary conditions to avoid edge effects. These are encoded in the $n \times n$ ``physical" inverse distance matrix $D_p$, where:

\begin{equation}
    D_{p, ij}  = 
    \begin{cases}
        c & \text{if} \;d(i,j) < 1/c \\
        1 / d(i,j) & \text{if} \;1/c < d(i,j) < r \\
        0 & \text{otherwise} \; .
    \end{cases}
\end{equation}
Here $d(i,j)$ denotes the Euclidean distance in physical (comoving) space between galaxies $i$ and $j$. To avoid approaching infinity in the limit of very small $d(i,j)$, a maximum cap $c$ can be imposed on $1/d(i,j)$. $D_p$ can be naively populated by calculating Euclidean distances between all pairs and then enforcing the neighborhood radius, setting all other elements to zero. It can also be implemented using a tree-based method such as KDTree \cite{kdtree} to only calculate necessary pairwise distances yielding a runtime more efficient than $\mathcal{O}(n^2)$. 

We then sort $D_p$ along each row from largest to smallest, such that now each row $D_{p,i}$ contains the sorted vector of inverse distances from galaxy $i$ to every other galaxy in the dataset. Since $r$ is much smaller than the box size, this vector likely contains many trailing zeros, as the vast majority of galaxy pairs are more than $r$ away from each other. We then calculate the Euclidean distances between each pair of sorted $n$-dimensional vectors of inverse distances to yield a second $n \times n$ matrix $D_s$, denoting ``similarity" distances between galaxy neighborhoods.

We use the sorted vectors of inverse distances to neighboring galaxies as our summary statistic of the geometry of a galaxy's local environment. The Euclidean distance between these sorted vectors (hereafter referred to as the ``similarity distance") yields a measure of how similar two galaxies' surrounding neighborhoods are. Galaxies with a very similar distribution of inverse distances to neighbors will have a low similarity distance, and galaxies with very different distributions will have a larger similarity distance. For two galaxies with different numbers of neighbors within the radius $r$, the excess nonzero distances will be compared against zeros, which will increase the similarity distance as expected, to denote that the two galaxies have different neighborhoods.

We then apply multidimensional scaling \cite[MDS;][]{Kruskal_1964}, a dimensionality reduction method that takes in a symmetrix matrix of pairwise distances and encodes them in a low-dimensional embedding space such that the pairwise distances are conserved as much as possible. MDS is remiscient of PCA \cite{pca}, its main difference being that it operates on a pairwise distance matrix instead of a covariance matrix. Here our similarity distance matrix $D_s$ serves as the input to MDS, with the first step being to ``center" it about zero:
\begin{equation}
    B = -\frac{1}{2} C D_s^2 C
\end{equation}
where 
\begin{equation}
    C = \mathbb{I} - \frac{1}{n}\mathbb{Q}_n
\end{equation}
where $\mathbb{I}$ is the identity matrix, $n$ is the number of points and $\mathbb{Q}_n$ is an $n \times n$ matrix of ones. 

From there, we take the eigendecomposition of the centered distance matrix $B$, yielding the matrix of eigenvectors $E$ and a diagonal matrix of eigenvalues $\Lambda$, and then scale the top $m$ eigenvectors by the square-root of their respective eigenvalues to yield the $m$-dimensional embedding space:
\begin{equation}
    X = E \Lambda^{1/2}
\end{equation}

Similar to PCA, the resulting embedding dimensions are ordered by how much information they contain- the top embedding dimensions preserves most of the similarity distances, and each following dimension preserves more of the similarity distances than the next. Thus, by keeping the top $m$ eigenvectors, we are creating the embedding space from the $m$ most informative dimensions.

Our method is similar to Isomap \cite{isomap}, which uses a graph representation to approximate geodesic distances between points and then applies MDS to the resulting distance matrix. Here, we instead use our neighborhood similarity distance matrix as an approximation of the parameter space we are looking to encode.

\subsection{Aligning embeddings across datasets}

Much like other dimensionality reduction methods such as UMAP \cite{umap} or t-SNE \cite{tsne}, a low-dimensional representation is entirely unique to the dataset that it encodes, and a different dataset's representation can be entirely different. More importantly, the latent dimensions of one dataset encode different information than those of another dataset, and they cannot be compared. The TNG simulations are expressed in redshift ``snapshots", separate datasets of the location and properties of the galaxies in the box at each point in time. Using our method naively, two different snapshots of the TNG simulations would be encoded separately into their own unique embedding spaces. In order to compare across datasets, they must be represented in exactly the same embedding space.

To address this, a data point from one dataset can be represented into the latent space of a different dataset by projecting it onto the eigenvectors of that dataset. To compare across redshift snapshots, we project the galaxies and their neighborhoods of earlier snapshots into the $z = 0$ snapshot. 

We can first rewrite the symmetric centered matrix $B$ as $F_cF_c^T$, where $F_C$ is the matrix of ``centered" (mean zero) inverse distance profiles from the $z = 0$ frame. This can then be decomposed using singular-value decomposition (SVD) where $F_c = USV^T$:
\begin{align}
        B &= F_c F_c^T\\
        &= (USV^T) (USV^T)^T \nonumber\\
        &= U S^2 U^T \nonumber\\
        &= E \Lambda E^T \nonumber
\end{align}

Through SVD, $U$ and $V$ both form orthonormal bases, and thus $U^{-1} = U^T$, showing it is equivalent to an eigendecomposition. Thus, $S = \Lambda^{1/2}$, $U = E$, and $F_c = E \Lambda^{1/2} V^T$.

The first step of this projection is to calculate the sorted vectors of inverse distances to neighbors of the earlier snapshot (say, the $z = 1$ snapshot) as described in the previous section - imposing the neighborhood radius cutoff $r$, calculating a pairwise physical distance matrix $D_{p, z = 1}$, and then sorting it along one axis. In this framework, the same matrix $D_{p, z = 0}$ has already been calculated for the $z = 0$ dataset.

We then calculate a cross-dataset similarity distance matrix $D_{s, cross}$ comparing the neighborhoods of galaxies in both datasets. If the $z = 0$ snapshot has $n$ galaxies and the earlier snapshot has $m$ galaxies, then we define an $n \times m$ matrix $D_{s, cross}$ where $D_{s, cross, ij}$ denotes the similarity distance between galaxy $i$ in the z = 0 snapshot and galaxy $j$ in the $z = 1$ snapshot. This similarity distance $S_{ij}$ is calculated as the Euclidean distance between row $i$ of $D_{p, z = 0}$ and row $j$ of $D_{p, z = 1}$. If $n$ and $m$ are not the same, then the longer row must be truncated to smaller of $(n,m)$. The truncation should happen at the end of the row such that the entries removed are mostly, if not entirely, leading zeros and thus do not affect the resulting similarity distances.

This $D_{s, cross}$ is then ``centered" in the $z = 0$ space to become $B_{cross}$, which can be written as 
\begin{equation}
    B_{cross} = P_c F_c^T
\end{equation}
where $P_c$ is the matrix of inverse distance profiles from the new ($z = 1$) frame, centered around the old ($z = 0$ frame) means.

The resulting matrix is then projected it onto the eigenvectors that denote the $z = 0$ embedding space:

\begin{align}
    X_{new} &= B_{cross} E \Lambda^{-1/2} \\
    &= P_cF_c^T E\Lambda^{-1/2} \nonumber \\
    &= P_c V  \Lambda^{1/2} E^T E\Lambda^{-1/2}\nonumber\\
    &= P_c V \nonumber
\end{align}
The resulting matrix $X_{new}$ is the projected embedding coordinates of each galaxy $j$ in the $z = 1$ snapshot in the location in embedding space where galaxy $j$ (and its neighborhood in the $z = 1$ snapshot) would be if it was in the $z = 0$ distribution of galaxies. The identity $X_{new} = P_c V$ means that $V$ can be calculated once from the $z = 0$ snapshot, and the projected coordinates $X_{new}$ can be calculated directly using just the inverse distance profiles from the new frame, eliminating the need to instantiate the $n \times n$ matrices $D_{s, cross}$ and $B_{cross}$.

A key difference is that since galaxy $j$ in the $z = 1$ snapshot was not actually in the $z = 0$ snapshot, it and its neighborhood do not play a part in shaping the $z = 0$ eigenvectors. Through this method, galaxies across snapshots can be encoded in the same embedding space, enabling direct comparisons of distributions of embedding values across redshift snapshots.

\subsection{Assumptions of this method}
There are several key assumptions of this method that should be noted. The first is the use of only radial distances when calculating pairwise distances between galaxies. While this is computationally efficient, we lose the angular information associated with different galaxy structures. For example, a set of galaxies that are collinear would encode exactly the same as a set of galaxies sitting at the same radial separations but spread across many angles. If angular information were to be taken into account, then structures such as filaments would be highly collinear and easier to distinguish; however, we assume that the distributions of radial distances are sufficiently different to distinguish between different kinds of structures. The inclusion of angular information would lead to additional computational complexity compared to calculating radial pairwise distances, and we leave this application to future work.

When projecting points from different datasets into the embedding space of another dataset, the underlying assumption is that the embedding space of the base dataset (the one being projected into) can encode all of the information in the new dataset. If there is a new structure present only in the new dataset, then the eigenvectors of the base dataset cannot express it. While the new points can be projected into the base embedding space, information that cannot be expressed in it is lost. For example, if the base dataset has no clusters but the new dataset does, the projected points of the new datset into the base dataset would not convey that there are clusters in the dataset, because this is not expressable in the base eigenbasis. In light of this, we use the $z = 0$ snapshot as the base dataset, as it likely has the most complex structures, and project the galaxy neighborhoods from earlier snapshots into the z = 0 latent space. 

In a uniformly distributed point cloud, any point is more likely to have more neighbors at larger distances. This effect equally impacts every point, and would be reflected in there being more smaller values in the sorted inverse distance vectors. {\tt SONDE} calculates Euclidean distances between sorted vectors of inverse distances to denote relative neighborhood similarity, accounting for this effect across pairwise similarity distances.

\subsection{Implementation}

The TNG100 dataset is a suite of detailed magnetohydrodynamical galaxy evolution simulations in a three-dimensional 75 cMpc/$h$ box \cite{tng1, tng2, tng3, tng4, tng5}. These simulations include detailed characterizations of key components of galaxy formation and evolution, including stellar evolution, chemical enrichment, heating and cooling of gas, and galactic winds \citep{2018MNRAS.473.4077P}, and have enabled significant work on galaxy-environment interactions \cite[e.g.][]{2019MNRAS.489..339H, 2020A&A...641A.173G, Yu_2025, andrews, 2025MNRAS.543..934U, 2026ApJ...997..233R}. The cosmology used in the simulations, which determines the nature of the cosmic web, is from \citet{2016A&A...594A..13P} with $\Omega_m = 0.3089, \sigma_8 = 0.8159$, and $H_0 = 100h \:\rm{km}\: \rm{s}^{-1}\rm{Mpc}^{-1}$ where $h = 0.6774$. The evolution of galaxies are captured in 100 ``snapshots", simulation boxes containing galaxy and halo positions and properties as they are measured across cosmic time from $z = 20$ to the present day ($z = 0$).

We create and compare embeddings for the TNG snapshots at $z = 0, 1, 2, 3, 4,$ and 5, spanning 12.6 billion years of evolution. For each TNG snapshot, we select confirmed subhalos with at least 500 star particles. For the $z = 5$ snapshot, there are 2,501 subhalos that meet this threshold with this number increasing to 5,199, 9,758, 16,833, 23,497, and 27,691 at $z = 4, 3, 2, 1$ and 0, respectively. We impose a neighborhood radius $r$ of 2 Mpc/$h$, motivated by the approximate size of a large dark matter halo \cite{2002PhR...372....1C}, although we find that the relationship between embedding dimensions and physical properties are stable to different values of $r$, as further-out distances are weighted less. We additionally impose a cap $c = 10$ on the inverse distances, meaning that all distances within 0.1 Mpc/$h$ are weighted equally.  

There are 760, 1,042, 1,142, 1,159, 1,168, and 1,138 galaxies with empty neighborhoods within 2 Mpc/$h$ at $z = 5, 4, 3, 2, 1$, and 0, respectively. All galaxies with empty neighorhoods encode to the same point in embedding space, at the coordinate (-7.29, 2.82, -1.30) in the first three embedding dimensions. We exclude these galaxies from downstream analyses to prevent them from biasing results.

We use the merger trees from the TNG simulations to connect galaxies at each snapshot with their earlier progenitors to track their evolution in embedding space. There are 1,079 subhalos at $z = 0$ that have progenitors in each of the previous snapshots, 497 of which have non-empty neighborhoods in each snapshot. The galaxies in the earlier snapshots are projected into the embedding space of the $z = 0$ snapshot as discussed above, and all embedding values are presented in this space.

Our method is scalable in that it does not require all pairwise distances to be calculated (only those within the neighborhood radius), and it can encode an entire dataset in a single matrix operation. By implementing a KDTree \cite{kdtree} to calculate only the necessary distances between galaxies and never having to create the entire pairwise distance matrix, it took 15.2 seconds to encode all 27,691 subhalos from the $z = 0$ snapshot on an Apple M2 Macbook Pro. 

\section{Results}\label{sec:results}

\subsection{Embedding Compared to Standard Clustering}

To understand how each embedding dimension is related to clustering, we measure the 2-point correlation function (2PCF) of all galaxies in the $z=0$ snapshot. This standard clustering estimator, $\mathcal{\xi}$, counts galaxy pairs at some separation $s$ in excess of randomly distributed positions. Although this contains no angular or higher-order information, its scale dependence is commonly used for probing the nonlinear behavior of how galaxies reside within and around dark matter halos \cite{2002ApJ...575..587B}.

We measure the autocorrelation (cross-correlation with themselves) of all galaxies, as well as the cross-correlation between the top and bottom 20\% of values for each dimension with the entire sample. Measurements were made using the full periodic box in logarithmic bins out to 20 Mpc/$h$ with the correlation package \textsc{Corrfunc} via \textsc{pycorr} \cite{2020MNRAS.491.3022S}. Errors were estimated with jackknife resampling of 27 subregions.

\begin{figure}
    \centering
    \includegraphics[width=\linewidth]{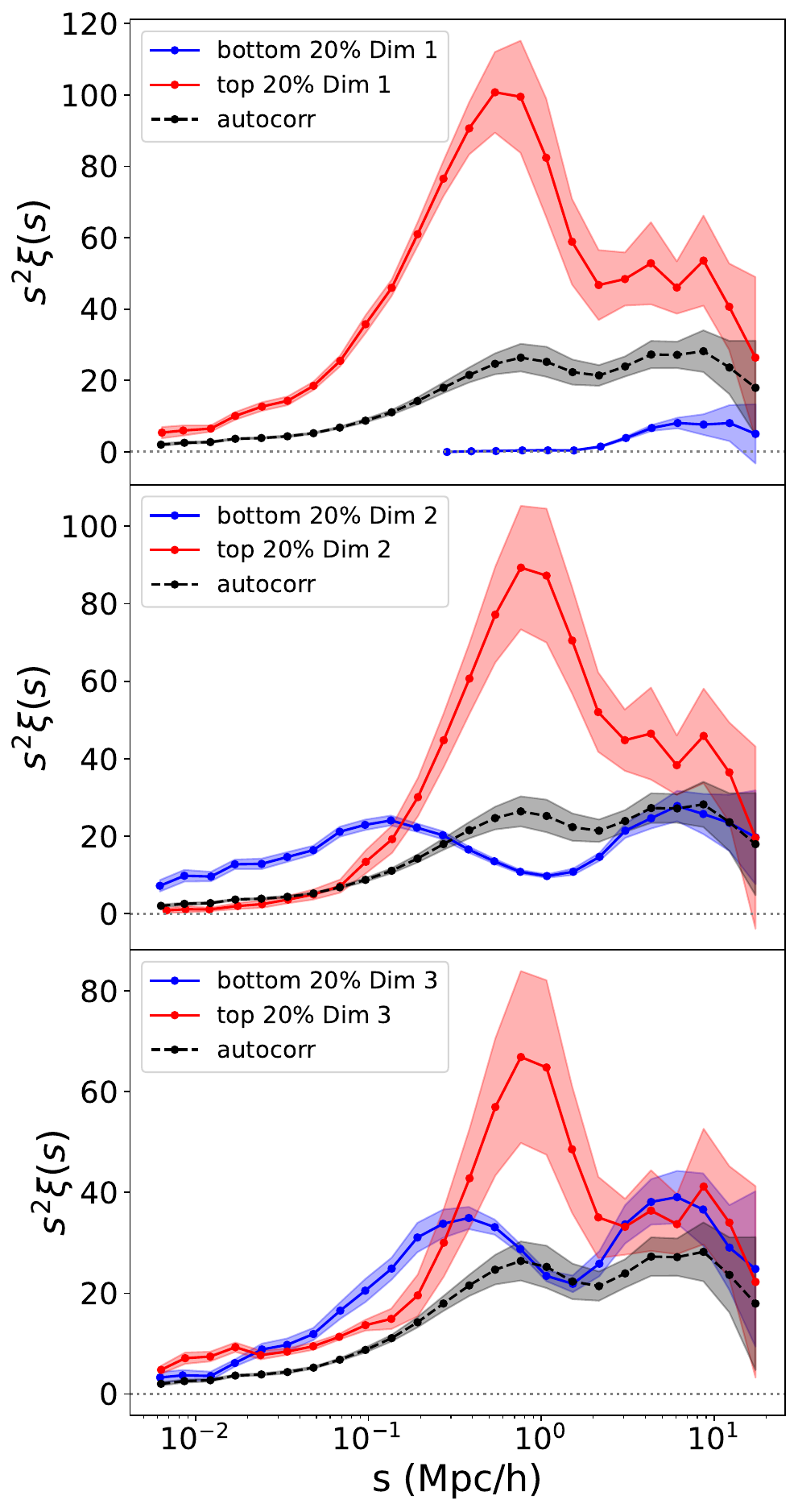}
    \caption{Galaxy cross-correlation functions between galaxies with the top and bottom 20\% values of the first (top), second (middle), and third (bottom) embedding dimensions with all galaxies in the $z = 0$ snapshot. Black dashed line denotes the autocorrelation function for all galaxies in the $z  = 0$ snapshot.}
    \label{fig:correlation_functions}
\end{figure}

Our results are displayed in Figure~\ref{fig:correlation_functions}. The autocorrelation is included in all subplots, and its dip around $s=2$ Mpc/$h$ corresponds to the one-to-two-halo transition \cite{Zehavi_2004}. Excess clustering arises at smaller scales from galaxies in common halos, followed by a flattening from halo exclusion. Correlations at the largest scales shown start to become limited by the finite periodic simulation volume, but do not impact qualitative comparisons between subsamples.

The first dimension shows the cleanest trend, as galaxies with the highest 20\% of Dim 1 values have the strongest clustering and the lowest 20\% have the weakest at all scales. This indicates that the first highest embedding dimension is largely driven by the same clustering that is quantified by standard clustering estimators, i.e. a higher Dim 1 value corresponds to galaxies in more clustered environments. There is additionally a lack of pairs at small separations when calculating the cross-correlation for the lowest values of Dim 1, which follows their being sparsely distributed.

The next two embedding dimensions show, on average, higher clustering with higher dimension values, but less distinctly and with different scale dependence. The clustering of galaxies with the lowest 20\% of Dim 2 values grows steeply above $s=2$ Mpc/$h$, indicating that it is likely capturing information related to the two-halo regime. 

Given the relatively small box size, cosmic variance may dominate overall trends at larger separations. The degeneracy in embedding space between dimensions also makes distinct interpretations of each dimension challenging. At minimum, the different scale dependence of 2PCF across embedding-defined subsamples demonstrates a sensitivity to information beyond standard clustering measurements. While the 2PCF alone cannot provide a comprehensive understanding of each embedding dimension, it lays a useful baseline for interpreting the correlations explored below and their connections to known trends.

\subsection{Distributions of distances to neighbors}

\begin{figure*}
        \centering
    \includegraphics[width=\linewidth]{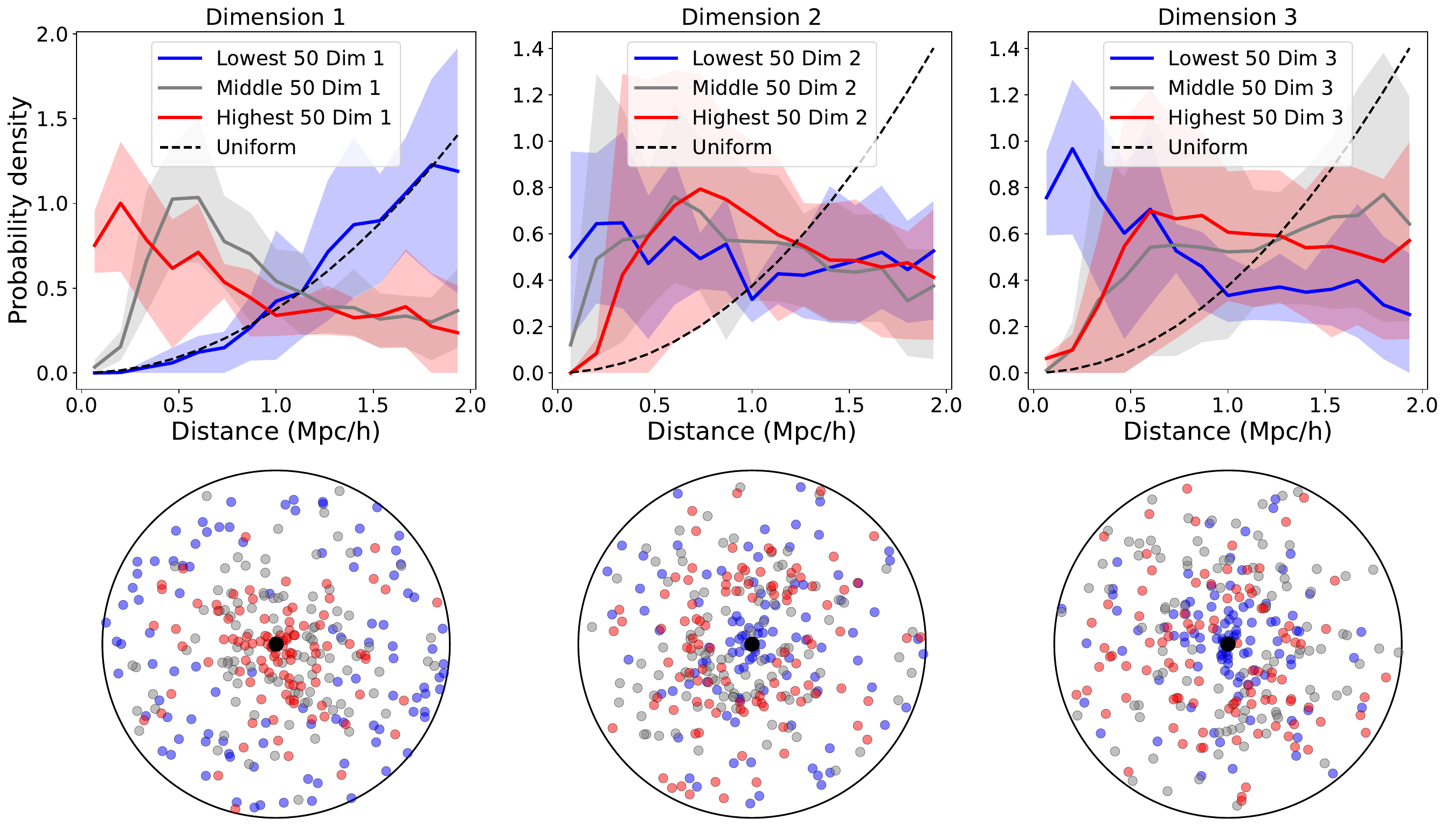}
    \caption{The distribution of distances to neighbors for galaxies with different values in embedding space. Normalized histograms (top row) and schematics (bottom row) of distributions of distance to neighbors for galaxies with 95-105 neighbors within 2 Mpc/$h$ in the $z$ = 0 snapshot. Each subplot shows the mean distribution for the 50 galaxies with lowest (blue), middle (gray), and highest (red) values of embedding dimension 1 (left), 2 (middle), and 3 (right). The solid line and  shaded region in the top row indicates the mean distribution and 16th-84th percentile band. Dashed line indicates the distribution of distances to neighbors from a uniform distribution of galaxy positions. Angles in the schematics below are randomly sampled for visualization purposes. These schematics qualitatively demonstrate the same information as above: how the values of each dimension correspond to different radial distributions of neighbors.}
    \label{fig:distributions}
\end{figure*}

{\tt SONDE} characterizes galaxy environments in terms of the distributions of inverse distances to their neighbors within a radius of 2 Mpc/$h$. When attempting to interpret the embedding space, we explore each of the top three embedding dimensions in the context of the distribution of distances to neighbors. 

The top row of Figure~\ref{fig:distributions} shows the normalized histogram of distances to neighbors for galaxies with the same number of neighbors (95-105) within the radius (2 Mpc/$h$) and their embedding values. Within each set of neighbors, we select the top, middle, and bottom 50 galaxies based on their value of each embedding dimension, and plot the mean and 16-84th percentile band of their distributions of distances to neighbors. This row thus shows how neighbors are distributed differently depending on the galaxy's location in embedding space. The bottom row shows schematics of the actual distribution of distances of 100 points, distributed at randomly sampled angles for visualization purposes. 

Figure~\ref{fig:distributions} demonstrates that even when the total number of neighbors within 2 Mpc/$h$ is the same, the embedding dimensions encode information about how these neighbors are distributed in this region. For example, for a constant number of neighbors, a high value (red line/points) of the first embedding dimension corresponds to more neighbors being distributed closer to the target galaxy. A low value of the first embedding dimension corresponds to a less clustered distribution of galaxies and aligns closely to a uniform distribution. Similarly, the second and third embedding dimensions seem to span a similar gradient of shapes of distributions including closely clustered and flat distributions.

The relationships between distribution shape and embedding values are not monotonic, and some of the embedding dimensions appear degenerate- for example, the middle and right panels look somewhat similar, with high values of both the second and third embedding dimensions corresponding to a similar distribution shape. This suggests that the two embedding dimensions may contain some of the same information. More distributions of distances to neighbors and their interpretations can be found in Appendix~\ref{appendix:distributions}.

\subsection{Correlation coefficents with interpretable statistics}

The embedding dimensions from {\tt SONDE} are determined from the distribution of inverse distances to neighbors. To understand the distribution characteristics represented in each embedding dimension, we calculate correlation coefficients between the top three embedding dimensions and various statistics of the distribution of distances to neighbors. These include the total number of neighbors within the neighborhood radius (2 Mpc/$h$), which is a proxy for local density. The median, minimum, and maximum distances provide summary statistics of the distribution of distances to neighbors. Finally, the standard deviation denotes the overall spread of the distribution, the skewness shows its asymmetry, and the kurtosis indicates sharpness of the distribution peak. Note that the the standard deviation is calculated only for galaxies with at least two neighbors, the skew and kurtosis only for galaxies with at least three neighbors, and all other correlation coefficients for galaxies with at least one neighbor within 2Mpc/$h$. The minimum neighbor distance does not impose the cutoff of 0.1 Mpc and is the minimum nonzero distance to a neighbor. Correlations between the top embedding dimensions and these statistics help interpret the information encoded in embedding space.

The resulting correlation coefficients are shown in Table~\ref{tab:correlation_table}.  The strong correlation between the first embedding dimension and the number of neighbors within the neighborhood radius solidifies that the first embedding dimensions can serve as a proxy for local density, consistent with the comparisons to standard clustering discussed above. The fact that this comes into the first embedding dimension suggests that local density is likely the most important interpretable statistic in characterizing a galaxy's surrounding environment. The lack of strong correlation of the third embedding dimension with any known statistics of the distributions suggests that it may encode higher-order information that is more difficult to interpret. Note that these correlation coefficients are calculated across the entire $z = 0$ galaxy population, but may be different when stratifying for galaxies with specific properties. Some of the coefficients may be positive or negative in different regimes, but average out to zero across the whole population.

\begin{table}[t]
    \centering
    \begin{tabular}{|c|ccc|}
    \hline
Statistic & Dim 1 & Dim 2 & Dim 3 \\
\hline
\# neighbors & \textbf{0.738} & \textbf{0.461} & 0.317 \\
Mean distance & \textbf{-0.473} & 0.346 & -0.091 \\
Standard deviation & 0.274 & \textbf{-0.484} & 0.222 \\
Skewness & 0.380 & -0.040 & -0.202 \\
Kurtosis & 0.067 & 0.062 & -0.011 \\
Minimum neighbor distance & \textbf{-0.544} & \textbf{0.414} & -0.317 \\
Median neighbor distance & \textbf{-0.469} & 0.283 & 0.000 \\
Maximum neighbor distance & 0.282 & 0.041 & -0.087 \\
\hline
    \end{tabular}
    \caption{Pearson correlation coefficients of each embedding dimension with various statistics of the distribution of distances to neighbors within 2 Mpc/$h$, excluding galaxies with empty neighborhoods. Correlation coefficients with magnitudes above 0.4 are bolded.}
    \label{tab:correlation_table}
\end{table}

\subsection{Embedding space and physical properties across redshifts}

\begin{figure}
    \centering
    \includegraphics[width=\linewidth]{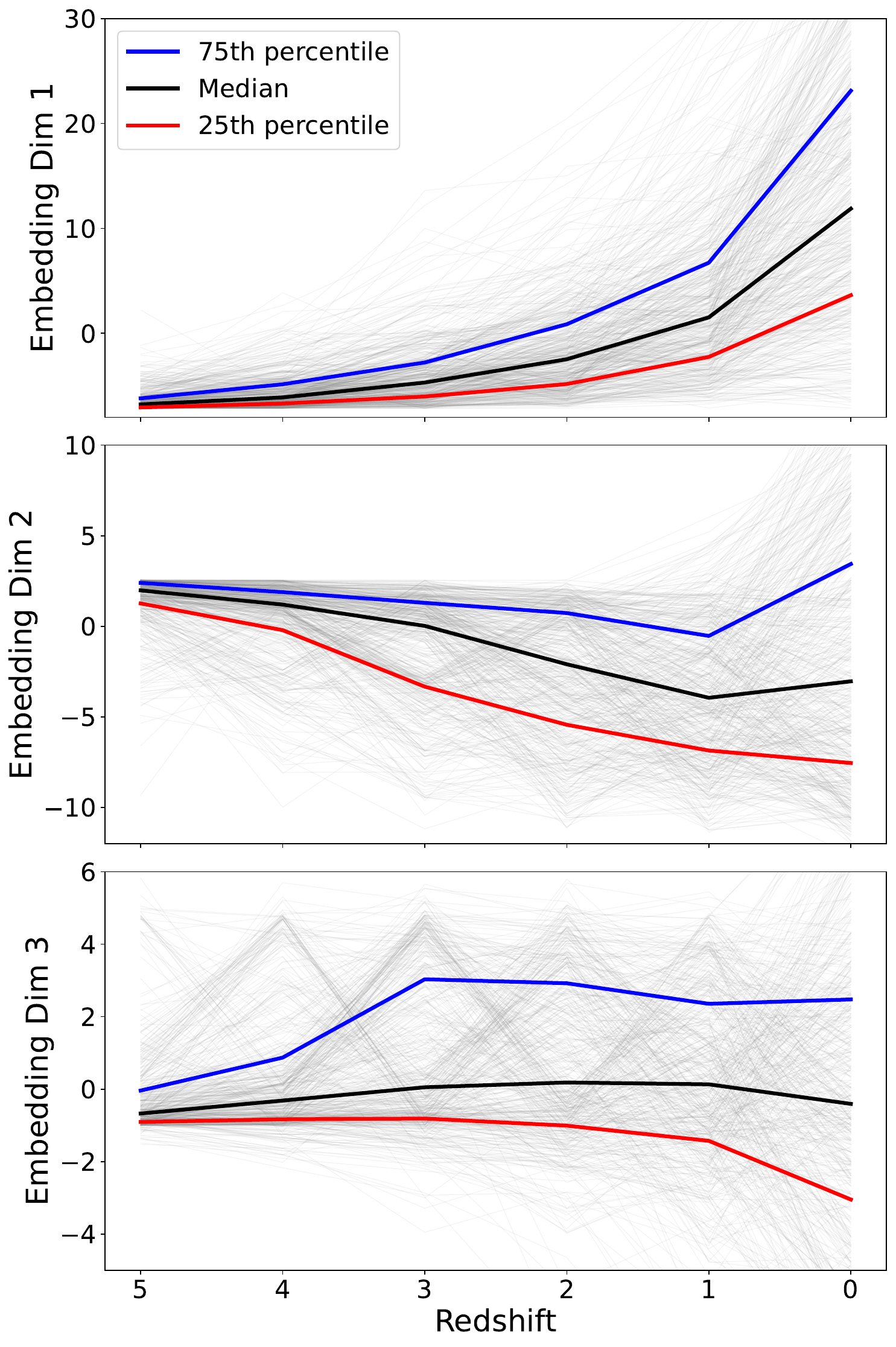}
    \caption{Values of the first, second, and third embedding dimensions (top, middle, and bottom panels, respectively) as a function of redshift for 497 galaxies in TNG100 with progenitors and nonempty neighborhoods at $z = 0 - 5$. Each thin gray line denotes the trajectory of one galaxy, and the blue, black, and red lines denote the 75th percentile, median, and 25th percentile of the population, respectively. Time increases across the $x$-axis, with redshift decreasing from $z = 5$ to our present-day Universe at $z=0$.}
    \label{fig:embedding_evolution}
\end{figure}

\begin{figure*}
    \centering
    \includegraphics[width=\linewidth]{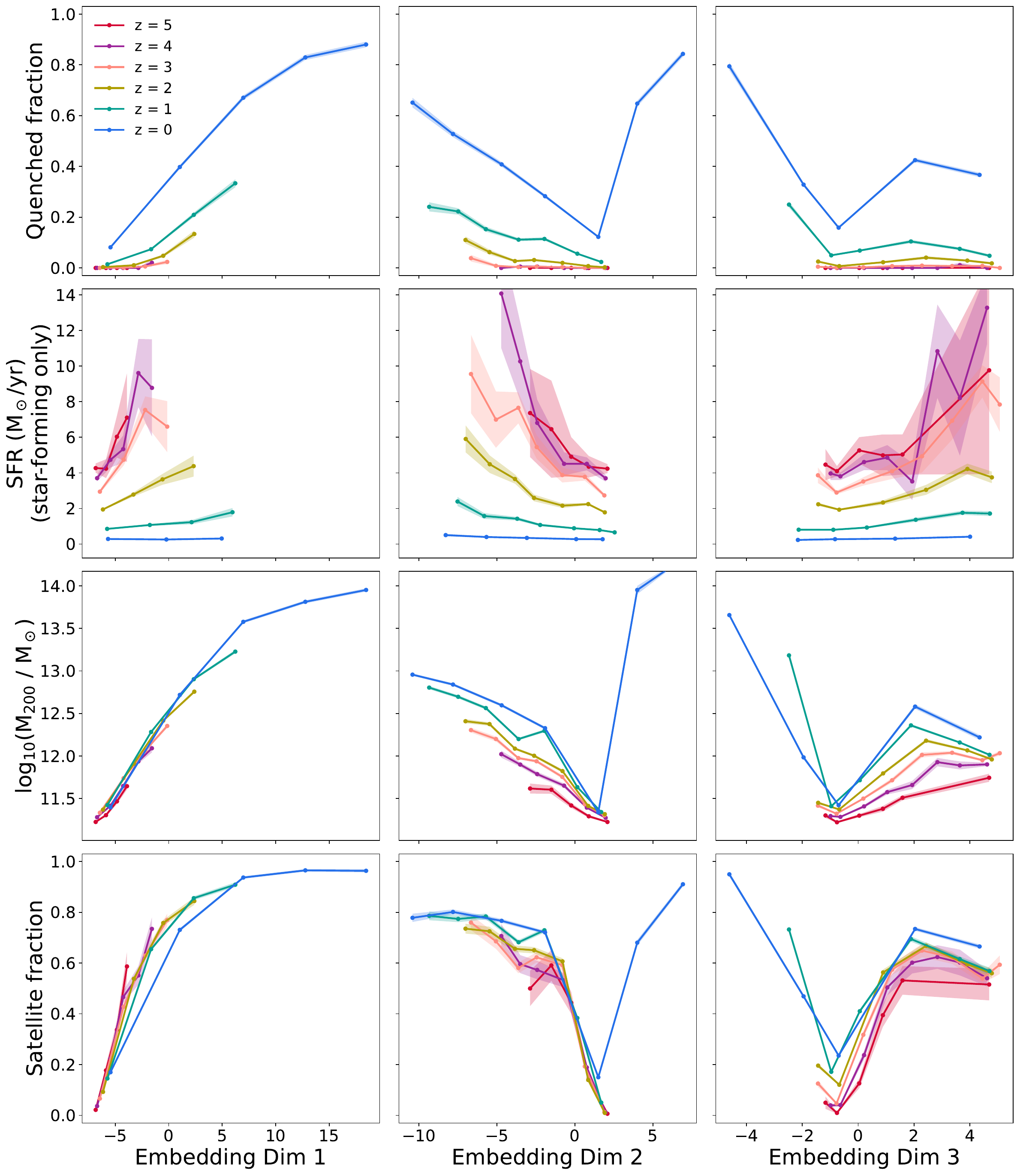}
    \caption{Quenched fraction (top row), star formation rate (second row), halo mass (third row), and satellite fraction (bottom row) as a function of the top three embedding dimensions (each column) for the $z = 0-5$ TNG100 snapshots. Lines and shaded regions denote the running median $\pm$ standard error (standard deviation/$n$) for the SFR and halo mass and the average fraction $\pm$ standard error (binomial confidence interval/$n$) for quenched and satellite fraction in each bin across all galaxies with non-empty neighborhoods in each snapshot. Bins containing less than 2\% of the snapshot's total number of galaxies with non-empty neighborhoods are not shown.}
    \label{fig:params_evolution}
\end{figure*}

To probe how galaxies' neighborhood geometries change over time, we track the embedding values of galaxies and their progenitors across redshifts. Figure~\ref{fig:embedding_evolution} shows the evolution in embedding space over redshift for each of the top three embedding dimensions. The top panel shows that galaxies monotonically increase in the value of their first embedding dimension, though the amount of change varies with each galaxy. The ranges of values of the first embedding dimension are different for each snapshot, with the earlier snapshots concentrated at the lowest values. The galaxy environments at earlier snapshots are projected into the $z = 0$ embedding space, so this evolution suggests that galaxies become more clustered over time. This follows the principle of hierarchical structure formation, where smaller structures form and then merge into larger, more complex structures \cite{1980PhyS...21..720P, 1984ApJ...284L...9K, 2005Natur.435..629S}. While the second and third embedding dimensions are harder to interpret, we see trends over redshift for both of them, with embedding values changing over time, suggesting that they encode some evolving aspect of large-scale structure. Furthermore, the spread of values in all dimensions increases with time, suggesting that environments also become more differentiated.

Figure~\ref{fig:params_evolution} shows the relationship between different physical parameters (quenched fraction, SFR, halo mass, and satellite fraction), plotted in one-dimensional bins of each embedding dimension. Here we impose a cutoff of specific SFR (SFR/galaxy mass) of $10^{-11}$ yr$^{-1}$, with galaxies below this threshold being classified as quenched, i.e. no longer star-forming. The second row shows SFR only for galaxies that are still star forming (with sSFR above this threshold). Satellite and quenched fraction are defined as the fraction of galaxies in each bin that are quenched vs. star-forming and satellite vs. central. We see the expected trends of physical properties with redshift- more galaxies are quenched, SFR decreases, and halo mass and satellite fraction increase over time \cite{2024ARNPS..74..173P}. 

Figure~\ref{fig:params_evolution} shows significant correlations between each embedding dimension and all of the physical properties in each redshift snapshot. This suggests that the embedding dimensions are encoding real physics that is directly related to the galaxies' physical properties. As {\tt SONDE} uses nonlinear dimensionality reduction, the exact nature of the functional mapping between the embedding dimensions and the physical properties is likely complex and difficult to interpret, but the observed trends indicate that the embedding contains important information nonetheless.

\begin{table*}[t]
    \centering
\centering
\caption{Results of an MLP trained to predict physical properties galaxies in the $z = 0$ snapshot from different combinations of features. Numbers shown are mean $\pm$ standard deviation from five-fold cross-validation. Classification tasks are scored as the AUC score and regression tasks are scored as $R^2$.}
\begin{tabular}{|c|cccc|}
\hline
Feature(s) & $\log M_{200}$ ($R^2$) & Satellite (AUC) & Quenched (AUC) & $\log \mathrm{SFR}$ ($R^2$) \\
\hline
$N_{\mathrm{nbr}}$ & $0.567 \pm 0.012$ & $0.826 \pm 0.008$ & $0.837 \pm 0.004$ & $0.003 \pm 0.001$ \\
Dim 1 & $0.760 \pm 0.004$ & $0.927 \pm 0.003$ & $0.880 \pm 0.005$ & $0.015 \pm 0.005$ \\
Dims 1--3 & $0.843 \pm 0.004$ & $0.959 \pm 0.002$ & $0.883 \pm 0.005$ & $0.042 \pm 0.006$ \\
Dims 1--10 & $0.842 \pm 0.004$ & $0.957 \pm 0.004$ & $0.883 \pm 0.005$ & $0.023 \pm 0.003$ \\
$N_{\mathrm{nbr}}$ + Dims 1--10 & $0.844 \pm 0.004$ & $0.956 \pm 0.004$ & $0.882 \pm 0.005$ & $0.023 \pm 0.007$ \\
$N(<r)$, 10 radii & $0.844 \pm 0.006$ & $0.957 \pm 0.002$ & $0.883 \pm 0.004$ & $0.030 \pm 0.006$ \\
\hline
\end{tabular}
\label{tab:embedding_prediction}

\end{table*}

\subsection{Predicting galaxy properties from embedding values}

To probe how informative our embedding space is compared to common metrics, we predict physical properties of each galaxy based on different combinations of these features. For this analysis, we use only the galaxies in the $z = 0$ snapshot as this is the largest individual dataset, and remove galaxies with empty neighborhoods, leaving 26,553 galaxies. The SFR is predicted only for 18,757 star-forming galaxies with sSFR $> 1 \times 10^{-11}\: \rm{yr}^{-1}$, as in the above section. We use a simple multilayer perceptron (MLP) neural network with two hidden layers to predict whether a galaxy is quenched or star-forming, whether it is a central or a satellite, its halo mass and SFR for star-forming galaxies.  An MLP with the same architecture is trained separately for each set of features and task (minimizing cross-entropy loss for classification tasks and mean-squared error for regression tasks), and we use five-fold cross-validation to evaluate and score the model. We explore predicting galaxy properties based on various sets of features: the total number of neighbors within 2 Mpc/$h$, the first 1, 3, and 10 embedding dimensions, the number of neighbors appended to the first 10 embedding dimensions, and the number of neighbors within each of 10 linearly spaced radius intervals from 0 to 2 Mpc/$h$. The results are shown in Table~\ref{tab:embedding_prediction}, evaluated with Area Under the Curve (AUC) for classification tasks and $R^2$ for regression tasks. Both metrics are constrained to be between 0 and 1, with higher values indicating better performance.

The results indicate that our embedding space contains more nuanced information about galaxies' environment than the number of neighbors. The first embedding dimension outperforms the number of neighbors in predicting all four physical properties, suggesting that it contains more relevant information. Adding the second and third embedding dimensions further improves predictions across all properties, though the quenched classification remains almost the same. Training the model on the first ten embedding dimensions yields very similar performance to the top three dimensions, with slightly worse SFR prediction. This suggests that most of the important information is contained in the top three dimensions, which follows as the nature of MDS means that, similarly to PCA, the embedding dimensions are ordered by the amount of information they retain. The addition of the number of neighbors to the top ten embedding dimensions makes no difference in performance. We compare to the cumulative number of neighbors within ten linearly spaced radius intervals, and this performs equally as well as the top three or ten embedding dimensions. The model performance suggests that environmental information is quite powerful for predicting the halo mass and a galaxy's quenched or satellite status.

The top three embedding dimensions has the best performance of all feature sets for predicting the SFR of star-forming galaxies, but the performance of all feature sets is extremely poor. The model performance suggests that the SFR is determined by other galaxy properties, such as its age, chemical composition, and gas density, that could be independent of its proximity to other galaxies \cite{1998ApJ...498..541K}. Note that the MLP is only trained and evaluated on the $z = 0$ snapshot from TNG100. The second row of Figure~\ref{fig:params_evolution} suggests that there may be a stronger relationship between the embedding dimensions and SFR at higher redshifts. 

A necessary caveat is that this model is trained and assessed only on physical properties from the TNG100 simulations. The IllustrisTNG subgrid model is calibrated to a small set of global observables. Subtle, real-universe correlations that manifold learning is uniquely sensitive to may not be present in TNG100. It is possible that this model would have different performance when applied to real data. Nonetheless, this analysis provides a quantitative demonstration that galaxy halo mass, central vs.  satellite status, and quenching likely correlate with its environmental geometry even further than just the number of neighbors, but that the SFR of star-forming galaxies is likely almost independent of their environments.

\section{Discussion}\label{sec:discussion}

This method can be generalized to identify additional aspects of a galaxy's environment. An important feature in the cosmic web are  voids, regions of space where there are no galaxies. Voids have been previously identified using the gradient and Hessian of the density field \cite{contardo}, and by introducing uniformly distributed points into an existing field of galaxies \cite{2025RASTI...4...32F}. The latter approach could be used to identify voids using our method. In this case, the modified dataset including the new uniform simulated points should not be projected into the embedding space of the original dataset, because its eigenbasis will likely not be able to represent void points. Instead, the uniform points should be introduced and a new embedding space should be created from the modified dataset. The new points that cluster in a new area of latent space, away from existing galaxies, likely denote void regions and indicate potential locations for follow-up. 

Our use of MDS provides an interesting comparison to contrastive learning, which has been used previously in astronomy to create informed latent representations \cite{2023RASTI...2..441H}. Using a contrastive loss function ensures that similar data points are closer together in latent space and different points are further apart, and is often used in conjunction with other loss functions (such as reconstruction loss) to structure the latent space of a variational autoencoder or other supervised deep learning models. Contrastive loss often uses predetermined sets of ``positive" or ``negative" pairs, denoting pairs of data points that have the same or different classifications, and adjusts the distances between pairs in latent space such that positive pairs are closer together and negative pairs are farther apart. MDS does this by design if a similarity distance matrix is used, and does not require a labeled training set to assign similarities. MDS only yields a linear decomposition based on the distance matrix, while using contrastive loss in conjunction with a supervised deep learning framework can yield a more complex latent representation of the data. Additionally, contrastive loss generally involves one or more hyperparameters that can control the exact level of clustering and additional regularization terms.

In this work, we calculate physical distances between galaxies in 3D comoving space, but this can be generalized to include other physical properties. The IllustrisTNG simulation suite contains many more properties of each galaxy that just its position. A 6D ``distance" could be calculated in joint position and velocity space, and this would result in galaxies clustering in latent space based not only on their physical proximity to other galaxies, but also how similarly they are moving. This could be a particularly useful application for identifying stellar streams and other structures characterized by their movement in addition to their shape. The distance metric could also be modified to include momentum and other relevant quantities depending on the use case. We could additionally explore distance metrics other than Euclidean to better structure the embedding space.

Currently, our method calculates similarity distances across each galaxy's sorted vector of inverse distances to neighbors. This could be further accelerated by expressing the neighbor distribution through a series of quantiles instead of passing in the entire distribution itself, although this presents an obvious trade-off with performance. If a reasonably large number of quantiles were used, much of the information about the shape of the distribution would be preserved and the latent space would not be too affected, while if too few quantiles were used, the method would essentially collapse to using known metrics such as the median distance and local density.

The scalability and expressivity of this method indicate that it could be a promising application for large observational datasets of 3D galaxy positions, but the use of real data adds the additional complications of completeness and redshift uncertainties. If a galaxy is not observed due to instrumental or systematic effects, the neighborhood of galaxies around it will not be accurately characterized. Redshift uncertainties and redshift space distortions will appear as non-physical structures if unaccounted for. To account for these effects, neighborhood similarity distances could be calculated weighting each galaxy based on the inverse of the uncertainty on its position. The effects of redshift-space distortions will likely require separate handling of projected and line-of-sight separations. The ideal dataset for this application is one from the Dark Energy Spectroscopic Instrument (DESI), which has measured highly accurate spectroscopic redshifts of millions of galaxies and has a relatively well-characterized selection function \cite{desi_dr1}.

Additionally, these large cosmological surveys require simulated data in volumes that are too large to practically generate with hydrodynamic simulations. Many use semi-analytic models, halo occupation distribution methods, and subhalo abundance matching, which map galaxies and their properties onto N-body simulations with varying degrees of complexity. This mapping requires a model of how galaxies are correlated with the surrounding structure, and therefore real-universe analysis can be limited by the order of information used, i.e. galaxy-halo connection models typically condition only on a small set of halo properties rather than full structural environment. Approaches like ours are a step towards efficiently encoding complex geometrical information that can produce realistic mock data more suited for probing higher-order correlations in a scalable way.

\section{Conclusions}\label{sec:conclusions}
We present a {\tt SONDE}, a novel manifold learning method for expressing the complicated geometry of galaxies' local neighborhoods in a low-dimensional embedding space. Using the distribution of inverse distances to neighbors to represent galaxy environments, we calculate similarity distances between galaxy pairs that denote how similar or different their local neighborhoods are. This pairwise similarity distance matrix is then used with multidimensional scaling in a variation on Isomap to create a low-dimensional embedding space such that galaxies with similar neighborhood structures cluster together.

We show that this embedding space is informative and encodes real physical relationships between galaxies' environments, their physical properties, and their evolution over time. The top three embedding dimensions evolve with redshift, and show relationships with physical properties. We explore the interpretation of the top three embedding dimensions by calculating correlation coefficients with known quantities, examining the two-point correlation function, and visualizing the shape of the distributions of distances to neighbors spanned by each dimension. We find that the first embedding dimensions correlates significantly with standard galaxy clustering, and the lower dimensions encode nuanced information about the shape of the distribution of distances to neighbors and large-scale structure around the galaxy. 

To explore the relationship between galaxy environments, we train an MLP to predict central vs. satellite, quenched status, halo mass, and star formation rate for each galaxy based on common metritcs and our top embedding dimensions. We find that a model trained on the top embedding dimension from {\tt SONDE} outperforms a model trained on just the number of neighbors within 2 Mpc/$h$, and that a model trained on the top three embedding dimensions performs equally as well as one trained on 10 radial bins. Additionally, star formation rate is very poorly predicted by all models, suggesting that it is independent of environment, but other parameters are relatively well-predicted just using environmental features.

Our method investigates the use of manifold learning for astrophysics and provides a scalable, continuous, and nuanced alternative to traditional methods of characterizing galaxy environments such as binary cluster identification, discrete structure classification, and local density estimation. Future applications include incorporating angular information, encoding large observational datasets, and incorporating uncertainties in the embedding space.

\begin{acknowledgments}

This work makes use of the {\tt scikit-learn} \cite{scikit-learn}, {\tt scipy} \cite{2020SciPy-NMeth}, {\tt pycorr/corrfunc} \cite{2020MNRAS.491.3022S} and {\tt numpy} \cite{harris2020array} Python packages. All code used in this work is publicly available at \url{https://github.com/asmuzsoy/galaxy-manifolds}. Generative AI Tools (GitHub CoPilot and Claude Code) were used for brainstorming, writing and debugging analysis and plotting code, generating \LaTeX{} tables, and debugging \LaTeX{} code. AI-generated code was based on human-written code and was closely supervised by the authors. The code was debugged, and the results were validated manually. The authors take responsibility for everything presented in this work.

ASMU was supported by a National Science Foundation Graduate Research Fellowship and AstroAI and would
like to thank Vedant Chandra, Michelle Ntampaka, Douglas Finkbeiner, Cecilia Garaffo, Andrew Saydjari, and Ashley Villar for helpful
discussions and suggestions. MW was supported in part by NSF award CBET-2112085 and DMS-2406905,
an Aramont Fellowship for Emerging Science Research, an Alfred P. Sloan Research Fellowship
in Mathematics, and the AI2050 program at Schmidt Sciences (Grant G-25-69786). CL is supported by an NSF Postdoctoral Fellowship under award 2502789.

\end{acknowledgments}

\appendix

\section{Detailed distributions stratified by embedding values}\label{appendix:distributions}

\begin{figure*}
    \centering
    \includegraphics[width=\linewidth]{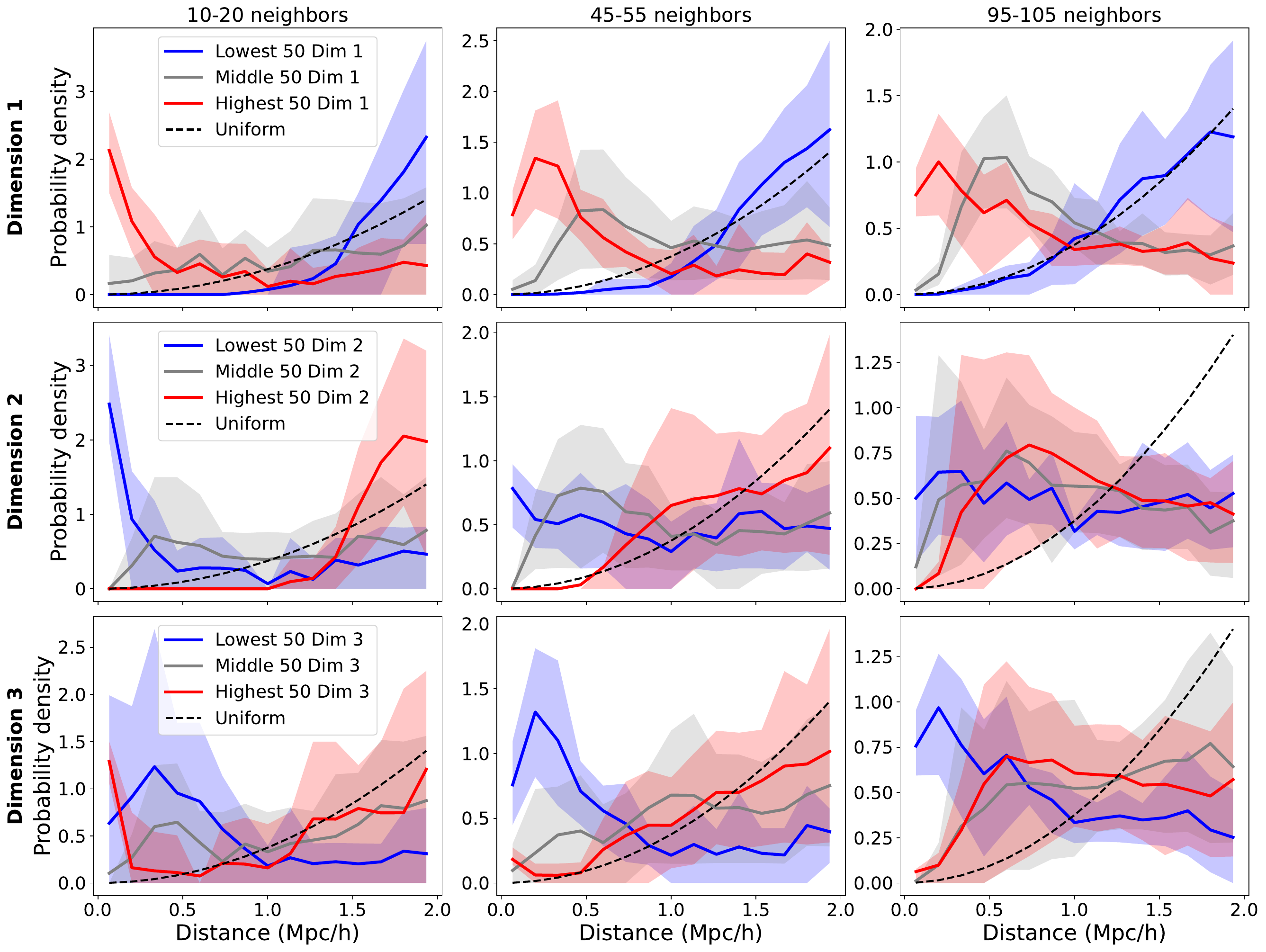}
    \caption{The distribution of distances to neighbors for galaxies with different values in embedding space. Shown are normalized histograms of distributions of distance to neighbors for galaxies in the $z$ = 0 snapshot. Each subplot shows the mean distribution for the 50 galaxies with lowest (blue), middle (gray), and highest (red) values of embedding dimension 1 (top row), 2 (middle row), and 3 (bottom row) with  10-20 (left column), 45-55 (center column), and 95-105 (right column) neighbors within 2 Mpc/$h$. The solid line denotes the mean distribution and the shaded region indicates the 16th-84th percentile band. Dashed line indicates the distribution of distances to neighbors from a uniform distribution of galaxy positions.}
    \label{fig:all_distributions}
\end{figure*}

The normalized distribution of distances to neighbors can have different shapes for different total numbers of neighbors within the neighborhood radius $r$ (here 2 Mpc/$h$). Figure~\ref{fig:all_distributions} is an extended version of the top row of Figure~\ref{fig:distributions}, showing the normalized histogram of distances to neighbors for galaxies stratified by number of neighbors within $r$ (either 10-20, 45-55, or 95-105) separated by their embedding values. Each column denotes a different total number of neighbors, and each row shows the normalized histogram stratified by a different embedding dimension. As an example, the left middle panel shows the distributions of distances to neighbors for galaxies at $z = 0$ that have 10-20 neighbors, with the distributions for the galaxies with the 50 lowest values of embedding dimension 2 in blue, the 50 middle values in gray, and the 50 highest values in red.

\bibliography{apssamp}

\end{document}